\documentclass[superscriptaddress,aps,journal=prl,preprint]{revtex4-2}
\usepackage{chemformula} 
\usepackage{graphicx}
\usepackage[inkscapelatex=false]{svg}
\usepackage{epstopdf}
\usepackage{bm,amsmath,amssymb} 
\usepackage{color, soul} 
\usepackage{dcolumn}   
\usepackage{multirow}
\usepackage{makecell}
\usepackage{booktabs}

\usepackage{hyperref}

\newcommand{\PB}{$\mathcal{P}_{\rm bg}$}
\newcommand{\PS}{$\mathcal{P}_{\rm s}$}
\newcommand{\PF}{$\mathcal{P}_{f}$}
\newcommand{\DEg}{$\Delta E_g$}
\begin{document}
\title{Sliding-Reversible Bandgap Modulation in Irreversible Asymmetric Multilayers}

\author{Changming Ke}
\affiliation{ Department of Physics, School of Science and Research Center for Industries of the Future, Westlake University, Hangzhou, Zhejiang 310030, China}
\affiliation{Institute of Natural Sciences, Westlake Institute for Advanced Study, Hangzhou, Zhejiang 310024, China}
\thanks{These three authors contributed equally}
\author{Yudi Yang}
\affiliation{ Department of Physics, School of Science and Research Center for Industries of the Future, Westlake University, Hangzhou, Zhejiang 310030, China}
\affiliation{Institute of Natural Sciences, Westlake Institute for Advanced Study, Hangzhou, Zhejiang 310024, China}
\thanks{These three authors contributed equally}
\author{Zhuang Qian}
\affiliation{ Department of Physics, School of Science and Research Center for Industries of the Future, Westlake University, Hangzhou, Zhejiang 310030, China}
\affiliation{Institute of Natural Sciences, Westlake Institute for Advanced Study, Hangzhou, Zhejiang 310024, China}
\author{Shi Liu}
\email{liushi@westlake.edu.cn}
\affiliation{ Department of Physics, School of Science and Research Center for Industries of the Future, Westlake University, Hangzhou, Zhejiang 310030, China}
\affiliation{Institute of Natural Sciences, Westlake Institute for Advanced Study, Hangzhou, Zhejiang 310024, China}

\date{\today}

\begin{abstract}
The electronic bandgap of a material is often fixed after fabrication. The capability to realize on-demand and non-volatile control over the bandgap will unlock exciting opportunities for adaptive devices with enhanced functionalities and efficiency. We introduce a general design principle for on-demand and non-volatile control of bandgap values, which utilizes reversible sliding-induced polarization driven by an external electric field to modulate the irreversible background polarization in asymmetric two-dimensional (2D) multilayers. The structural asymmetry can be conveniently achieved in homobilayers of Janus monolayers and heterobilayers of nonpolar monolayers, making the design principle applicable to a broad range of 2D materials. We demonstrate the versatility of this design principle using experimentally synthesized Janus metal dichalcogenide (TMD) multilayers as examples. Our first-principles calculations show that the bandgap modulation can reach up to 0.3 eV and even support a semimetal-to-semiconductor transition. By integrating a ferroelectric monolayer represented by 1T$'''$-MoS$_2$ into a bilayer, we show that the combination of intrinsic ferroelectricity and sliding ferroelectricity leads to multi-bandgap systems coupled to multi-step polarization switching. The sliding-reversible bandgap modulation offers an avenue to dynamically adjust the optical, thermal, and electronic properties of 2D materials through mechanical and electrical stimuli.

\end{abstract}

\maketitle
\newpage


The bandgap, a fundamental property of materials, is crucial in determining their optical, thermal, and electrical behaviors~\cite{Ashcroft76}. It holds a key role in modern device physics and technology, determining the performance of a wide range of devices like transistors, solar cells, and light-emitting diodes~\cite{Zse06Book,Chaves20p29,Varley22p230401,Ning17p17070}. The ability to dynamically and non-volatilely adjust the bandgap is highly desirable, enabling real-time control of material properties. This opens up opportunities for ``smart devices" that respond instantaneously to changes in environmental conditions or user inputs, thereby enhancing both functionality and efficiency~\cite{Zhou21p106613}. For example, solar panels with tunable bandgaps could optimize their absorption characteristics to maximize efficiency throughout the day and in various weather conditions. Materials with temperature-sensitive bandgaps that regulate light and heat transmission in response to changing environmental conditions form the basis of smart windows~\cite{Zhou21p106613}.
Similarly, tunable laser diodes that allow on-demand control of their output wavelength may provide enhanced precision in procedures like laser surgery or scanning~\cite{Buus05Book}. 
While techniques like doping~\cite{Chen13p4610,Ning17p17070}, strain~\cite{Lee04p011101}, and interface engineering~\cite{Kroemer01p783} can effectively modify material bandgaps, these methods are typically static, meaning the bandgap is fixed after fabrication. This poses a challenge for achieving dynamic and real-time control over bandgaps. Efforts for reversible bandgap tuning have mainly focused on phase-change materials~\cite{Dong19p1806181,Prabhathan23p107946} and electric field effect~\cite{Zhang09p820,Castro07p216802}. However, materials that exhibit robust, dynamic, and non-volatile bandgap tunability are still relatively rare.

Two-dimensional (2D) van der Waals (vdW) materials can be stacked and relatively shifted. The interlayer motions like sliding and twisting offer a unique degree of freedom to tune the properties of these material, even inducing  emergent phenomena not present in their individual layers~\cite{ViznerStern21p1462,Wu21pe2115703118,Cui20peabb1335,Pakdel24p932,Ding21p084405}.
For example, 
controlling the twist angle between layers can lead to the formation of long-wavelength moir\'{e} potential, effectively quenching the kinetic energy of electrons and enhance the electronic correlation. This sets the stage for various exotic quantum phases, such as superconductivity~\cite{Cao18p43, Chen19p215, Yankowitz19p1059} and Mott insulators~\cite{Chen19p237, Shimazaki20p472, Cao18p80}.
In recent years, 
sliding ferroelectricity, a phenomenon where lateral sliding between nonpolar monolayers produces a switchable vertical polarization, has attracted great attention. 
Notable bilayer systems exhibiting sliding ferroelectricity that have been confirmed experimentally include bilayer boron nitride and transition metal dichalcogenides~\cite{Yasuda21p1458, ViznerStern21p1462, Zheng20p71, Roge22p973, Wu11p236101}.
This discovery expands the range of 2D ferroelectrics with out-of-plane polarization, potentially allowing for full utilization of atomic thickness in the development of ultrathin nanoelectronics.
Various measures can induce interlayer sliding in layered vdW materials, including external electric field~\cite{Tsymbal21p1389,Yasuda21p1458,ViznerStern21p1462} and shear strain~\cite{Liu22p38}.

Given the facile capability to induce interlayer sliding, sliding-induced bandgap modulation, if achievable, could significantly improve the methods available for real-time tuning of electronic structures. However, unlike twisting, sliding in bilayers of identical monolayers generally has a less significant impact on their electronic properties, as it doesn't substantially modulate the periodic potential landscape.
Here, we present a general strategy that 
integrates a reversible polarization arsing from sliding (\PS) with a background out-of-plane polarization (\PB) 
that does not need to be reversible (switchable), to achieve sliding-tunable bandgaps in a vdW multilayer.

\begin{figure}[t]
\centering
\includegraphics[scale=0.58]{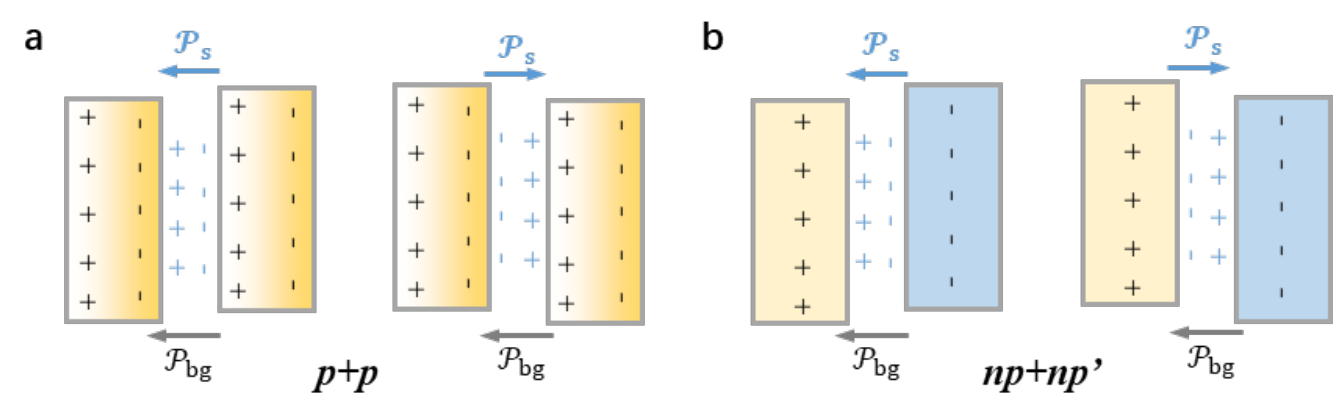}
\caption{
Design principle for sliding-reversible bandgap modulation in asymmetric bilayers. Here, {\em p} denotes a polar layer such as a Janus monolayer, {\em np} a nonpolar layer, and {\em np'} a different nonpolar layer. (a) $p+p$ stacking with the background polarization $\mathcal{P}_{\rm bg}$ arising from the intrinsic asymmetry of the monolayer. (b) $np+np'$  stacking with~\PB~resulting from the charge transfer due to the work funciton difference. The yellow layer is the low-work-function layer, and the blue layer is the high-work-function layer. In both cases, \PB~is not reversible, while \PS~is reversible by interlayer sliding.
}
\label{fig_sche}
\end{figure}

Taking a bilayer as an example (Fig.~\ref{fig_sche}), the background polarization may result from the structural asymmetry of the monolayer, such as in a homobilayer of Janus materials. It could also arise from charge transfer between nonpolar monolayers with different work functions in a heterobilayer. In both cases, the direction of \PB~is \textit{\textbf{not}} reversible by an electric field. 
Another key requirement for the bilayer configuration is that it should support sliding ferroelectricity, which can be conveniently achieved based on the general theory for bilayer stacking ferroelectricity~\cite{Ji23p146801}. 
For a bilayer with both \PB~and \PS, the sliding allows the transition between two states with different out-of-plane polarization, a high-polarization (HP) state with $\mathcal{P}_{\rm bg}-\mathcal{P}_{\rm s}$ and a low-polarization (LP) state with  $\mathcal{P}_{\rm bg}+\mathcal{P}_{\rm s}$. Since the depolarization field within the bilayer is directly coupled to the magnitude of the out-of-plane polarization and can strongly affect the band gap~\cite{Duan21p2316,Ke21p3387,Huang22p1440}, the two polar states can possess substantially different bandgaps, enabling sliding-induced, nonvolatile tuning of the electronic structures. 

We demonstrate this design principle using density functional theory (DFT) calculations, focusing on the well-known family of 2D materials, transition metal dichalcogenides (TMDs)~\cite{Manzeli17p17033,Voiry15p2702}, many of which have already been synthesized experimentally. We find that a Janus 1H-$AXY$ ($A$ = Mo, W; $X, Y$ = S, Se, Te) bilayer with parallel stacking, denoted as the $p+p$ configuration (Fig.~\ref{fig_sche}a), can exhibit a reversible bandgap change of 0.1--0.2 eV. Moreover, a sliding-driven semimetal-semiconductor transition is realized in the 1T-WTeSe bilayer. 
Additionally, the magnitude of bandgap modulation can be further enhanced with increasing layer number. Specifically, 
the sliding-induced bandgap change in the trilayer 1H-WTeSe reaches up to 0.3 eV. The design principle is also applicable to a heterobilayer consisting of nonpolar monolayers, denoted as $np+np'$ (Fig.~\ref{fig_sche}b), as exemplified by 1H-WSe$_2$/1H-Mo$X_2$ ($X$ = S, Se, Te). Finally, we show that 
in a bilayer consisting of recently synthesized ferroelectric 1T$'''$-MoS$_2$ and nonpolar 1H-MoS$_2$ ($f+np$ configuration), it is possible to achieve a multi-bandgap system coupled to multistep polarization switching. The design principle proposed in this work for on-demand control of bandgaps is versatile and opens up opportunities for voltage-configurable multi-state electronics.

DFT calculations are performed using the Vienna {\em ab initio} Simulation Package \texttt{VASP}~\cite{Kresse96p11169,Kresse96p15}.  The exchange-correlation interaction is described by the Perdew–Burke–Ernzerhof (PBE) functional~\cite{Perdew96p3865} and Heyd-Scuseria-Ernzerhof hybrid density functional (HSE06)~\cite{Krukau06p224106}, with Grimme's D3 dispersion correction~\cite{Grimme10p154104}. The projector augmented wave (PAW) method~\cite{Blochi94p17953} is used to describe the electron-ion interaction between the core ion and valence electrons. The plane-wave kinetic energy cutoff is 600~eV, and a 9$\times$9$\times$1 Monkhorst-Pack grid is used to sample the Brillouin zone for the unit cell. A vacuum region of more than 20~\AA~thick is included to avoid interactions between periodic images. The climbing image nudged elastic band (CL-NEB) method \cite{Henkelman00p9901} is employed to identify minimum energy pathways, with electronic energy and atomic force converged to 10$^{-7}$ eV and 0.01 eV/\AA, respectively. Dipole correction is considered in all calculations. 

\begin{figure}[t]
\centering
\includegraphics[width=16cm]{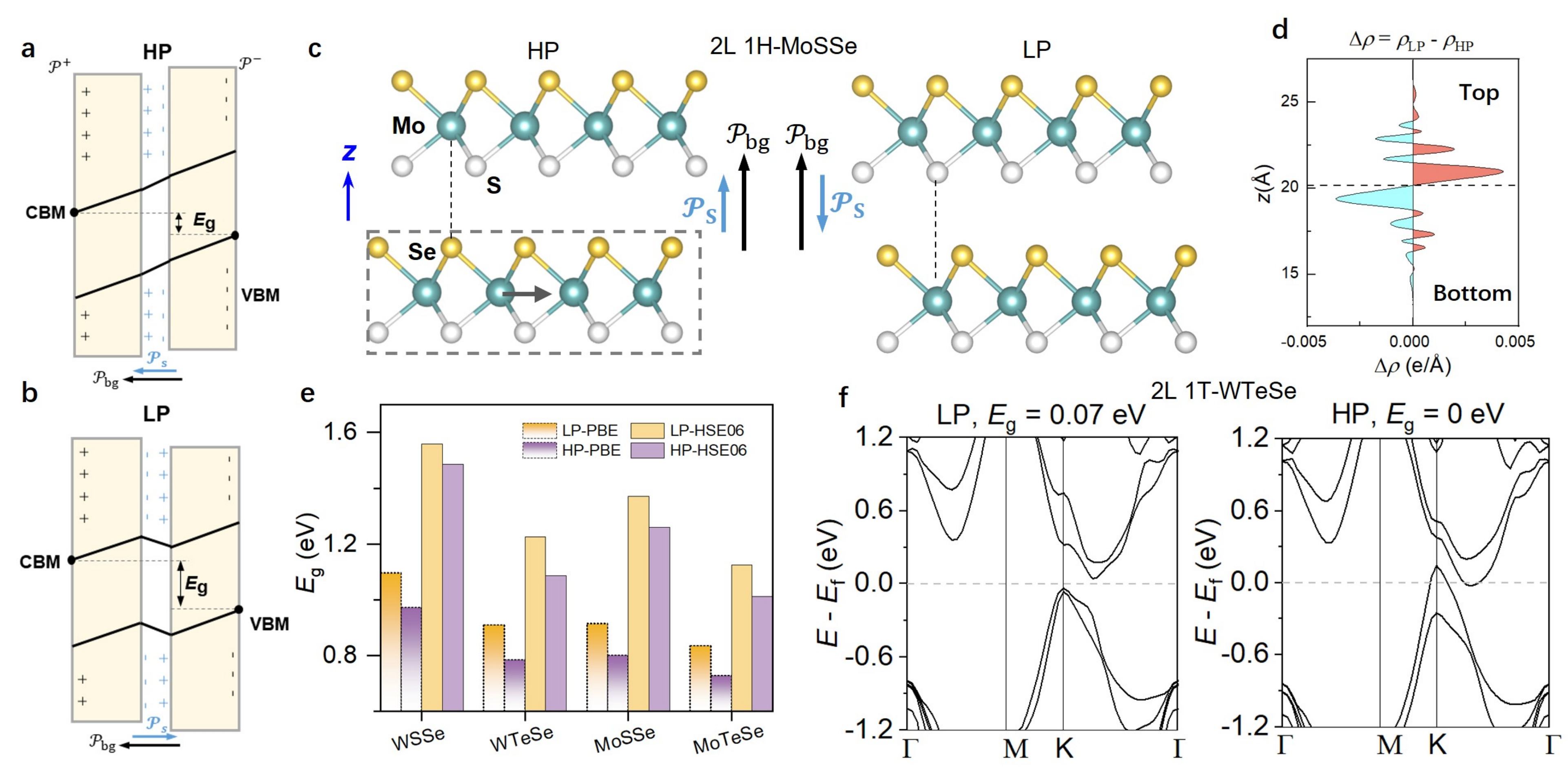}
\caption{Sliding-reversible bandgap modulation in Janus TMD homobilayers of $p$+$p$ configuration. Schematics of band alignments in (a) high-polarization (HP) and (b) low-polarization (LP) states. (c) Configurations of bilayer (2L) 1H-MoSSe showing sliding-induced flip of \PS~from +$z$ to $-z$. (d) Sliding-induced interlayer charge transfer along the $z$ direction, with $\Delta \rho$ representing the difference in the electronic charge density between LP and HP states ($\Delta \rho>0$ indicates an increase in electron density). (e) Bandgap values for bilayer 1H-WSSe, WTeSe, MoSSe, and MoTeSe, computed with PBE and HSE06. (f) Sliding-induced semiconductor-semimetal transition in the 2L 1T-WTeSe, with band structures computed using HSE06.}
\label{fig_polar_2L}
\end{figure}

We start with a heuristic approach to understand the bandgap modulation caused by the interlayer sliding in a homobilayer consisting of two polar Janus monolayers. As depicted in Fig.~\ref{fig_polar_2L}a, the surface polarization bound charges associated with the out-of-plane polarization generates a depolarization field that opposes the polarization. This field establishes a potential step that scales with the magnitude of the polarization, thereby differentiating the electronic states of the two polar surfaces. Electrons near the $\mathcal{P}^{-}$ surface naturally have higher energies due to increased Coulomb repulsion, which defines the valence band maximum (VBM). Conversely, the conduction band minimum (CBM) is located at the $\mathcal{P}^{+}$ surface. The depolarization field resembles the built-in electric field within the depletion region of a $p$–$n$ junction. Although the polarization direction of Janus monolayers, \PB, are not reversible, the polarization arsing from the stacking, \PS, can be flipped by sliding. Therefore, the effective potential step across the bilayer in the HP state with $\mathcal{P}_{\rm bg}+\mathcal{P}_{\rm s}$ is higher than that in the LP state with $\mathcal{P}_{\rm 
bg}-\mathcal{P}_{\rm s}$. This results in a larger bandgap in the LP state.


The simple band bending model discussed above is corroborated by our DFT calculations of 2D Janus TMDs, many of which, like MoSeS and WSeS, have already been synthesized~\cite{Lu17p744, Feuer23p7326}. As a case study, we focus on the 1H-MoSSe bilayer. In the AB-stacking configuration, as illustrated in Fig.~\ref{fig_polar_2L}c, electrons transfer from the upper Mo layer to the lower Se layer due to the greater electronegativity of Se relative to Mo. This leads to an upward \PS~that runs along \PB~and thus a HP state.
Band structure calculations with PBE reveal a bandgap of $E_g^{\rm HP}=1.48$~eV. Sliding the bottom layer by 1/3 of a unit cell reverses the charge transfer direction (see Fig.~\ref{fig_polar_2L}d), resulting in a LP state and a higher bandgap of $E_g^{\rm LP}=1.58$~eV. Therefore, a bandgap modulation $\Delta E_{\rm g}$ = 0.1 eV is achieved in the 1H-MoSSe bilayer. We further perform calculations for 1H-WSSe, WTeSe, MoSSe, and MoTeSe bilayers, with the bandgap values computed with PBE and HSE06 reported in  Fig.~\ref{fig_polar_2L}e. It is evident that the bandgap modulation driven by the sliding is a robust phenomenon common to all studied bilayers. 
Besides the 1H phase with trigonal prismatic coordination, 2D TMDs may also crystallize in the 1T phase featuring the tetragonal symmetry and octahedral coordination of the transition metal. 
Interestingly, we find that the bilayer 1T-WTeSe undergoes a semiconductor-to-semimetal transition during the sliding from a LP state to a HP state, as depicted in Fig.~\ref{fig_polar_2L}f.  

\begin{figure}[t]
\centering
\includegraphics[width=12cm]{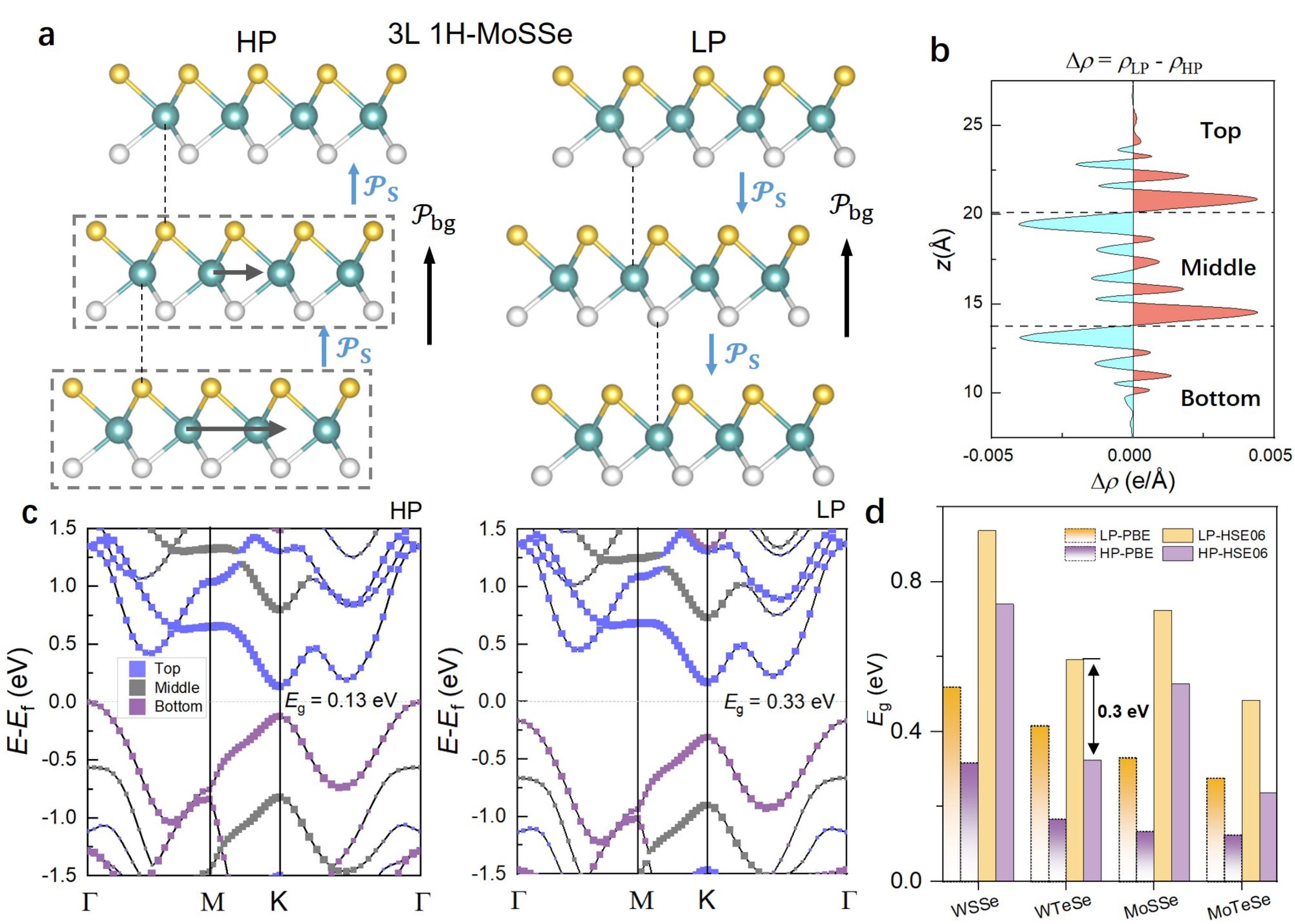}
\caption{Enhanced reversible bandgap modulation in Janus trilayers. (a) Configurations of trilayer (3L) 1H-MoSSe illustrating the transition from a ABC-stacking to CBA-stacking by a sliding gradient. The directions of \PS~at both interfaces are flipped. (b) Sliding-induced interlayer charge transfer in the $z$ direction. A positive $\Delta \rho$ value means an increase in electron density. (c) Atom-resolved band structures in 3L 1H-MoSSe. (d) Bandgap values for trilayer 1H-WSSe, WTeSe, MoSSe, and MoTeSe at different polarization states.}
\label{fig_polar_3L}
\end{figure}

The magnitude of \DEg~can be enhanced in multilayers by promoting the contribution of reversible \PS~to the total polarization.
For a trilayer 1H-MoSSe, it may adopt ``ladder ferroelectricity" where each pair of neighboring monolayers supports stacking ferroelectricity~\cite{Deb22p465}.  As illustrated in Fig.~\ref{fig_polar_3L}a, an ABC stacked trilayer has two interfaces possessing parallel upward \PS. Applying a sliding gradient, where the middle layer slides one-third of a unit cell and the bottom layer slides two-thirds of a unit cell, will change the stacking from ABC to CBA, switching the direction of \PS~at both interfaces and resulting in a LP state (see the change in charge density in Fig.~\ref{fig_polar_3L}b). 
Our DFT calculations confirm that the ABC configuration, characterized by an out-of-plane polarization of 18.6 pC/m, has a PBE band gap of 0.13~eV. In contrast, the polarization in the CBA-stacked trilayer decreases to 16.7 pC/m, and the bandgap widens to 0.33 eV. The magnitude of $\Delta E_{\rm g}$ in this trilayer is 0.2~eV, doubling the 0.1~eV observed in the bilayer. 
Additionally, the atom-resolved band structures (Fig.~\ref{fig_polar_3L}c) confirm that the VBM and CBM are dominated by states of the bottom $\mathcal{P}^-$ layer and the top $\mathcal{P}^+$ layer, respectively, consistent with the schematic illustrated in Fig.~\ref{fig_polar_2L}a and b. 
As summarized in Fig~\ref{fig_polar_3L}d for several Janus trilayers, the value of \DEg~ranges from 0.2 to 0.3~eV, improving upon those in bilayers. Specifically, HSE06 predicts a bandgap modulation of $\approx$0.3~eV and a direct-to-indirect bandgap transition in the trilayer 1H-WTeSe.

\begin{figure}[t]
\centering
\includegraphics[scale=0.40]{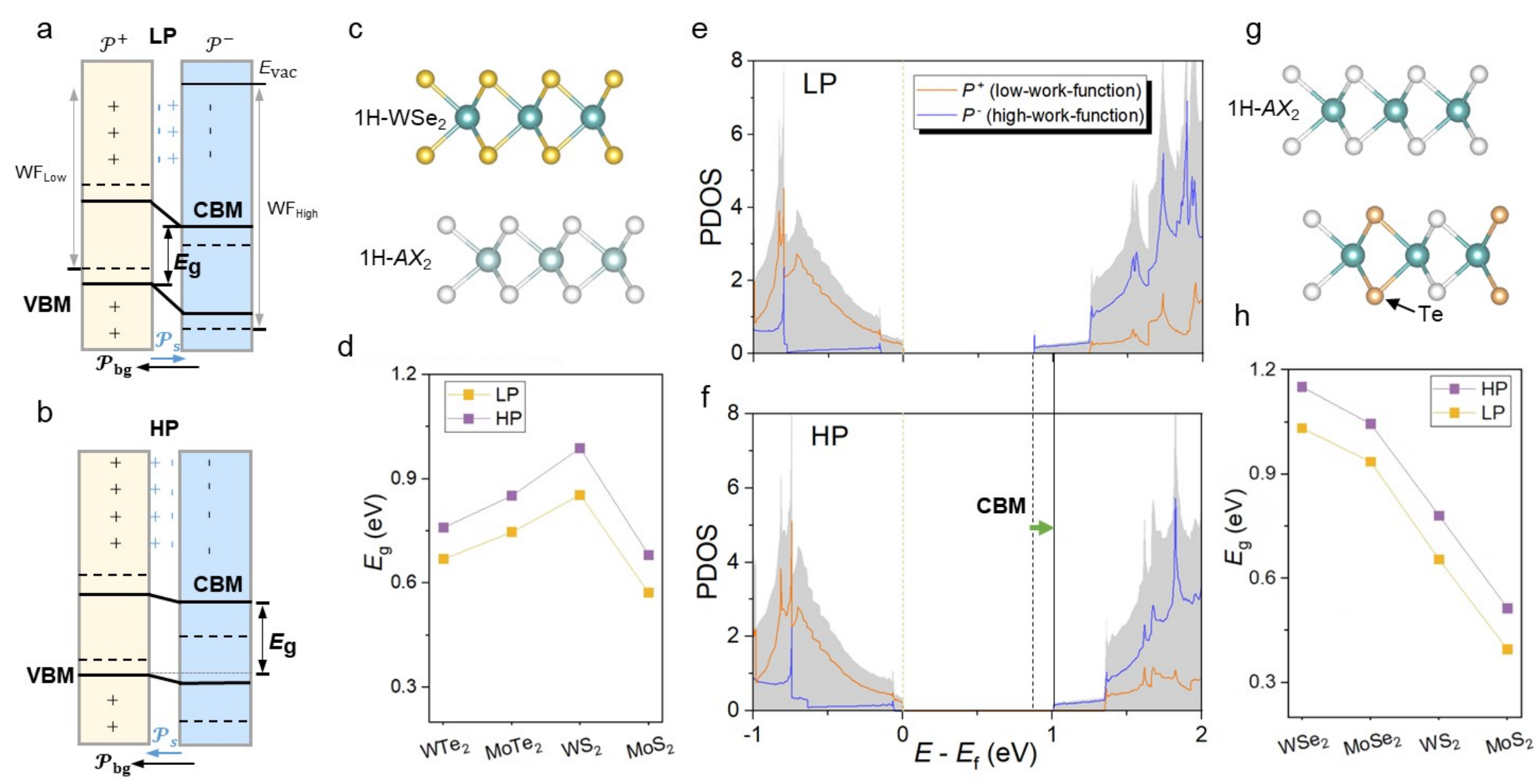}
\caption{
Sliding-reversible bandgap modulation in hetrobilayers consisting of two nonpolar monolayers ($np$+$np'$).
Schematics of band alignments in (a) low-polarization (LP) and (b) high-polarization (HP) states. The dashed lines represent the intrinsic CBM and VBM before charge transfer. The yellow and blue layers have low-work-function (WF$_{\rm Low}$) and high-work-function (WF$_{\rm High}$), respectively. (c) Atomic structure of 1H-WSe$_2$/1H-$AX_2$ ($A$ = Mo, W, $X$ = S, Te). (d) Bandgap values for 1H-WSe$_2$/1H-$AX_2$ bilayers in HP and LP states computed with PBE. Projected density of states (PDOS) of (e) LP  and (f) HP states for 1H-WSe$_2$/1H-WS$_2$ heterobilayer. (g) Atomic structure of an asymmetrically doped (quasi-)homobilayer consisting of 1H-$AX_2$ ($A$ = W, Mo; $X$ = S, Se) and its homogeneously Te-doped nonpolar monolayer 1H-$AX{\rm Te}$. The bandgap values are reported in (h).}
\label{fig_hetero}
\end{figure}

The proposed design principle does not rely on polar monolayers for two reasons. First, the work function difference between the constituent monolayers in a heterobilayer can serve as the origin of \PB. Second, a distinct feature of sliding ferroelectrics is that the stacking of nonpolar monolayers can also generate vertical polarization as long as the stacking satisfies the symmetry requirement~\cite{Ji23p146801}. The mechanism of bandgap modulation in an $np$+$np'$ heterobilayer is sketched in Fig.~\ref{fig_hetero}a and b, which differs subtly from the $p$+$p$ stacking. 
The charge transfer between monolayers pushes down the valence states in the low-work-function layer and raises the conduction states in the high-work-function layer. Since the VBM and CBM are located in the low-work-function and high-work-function layers, respectively, stronger charge transfer causes the VBM and CBM to move further apart. In the LP state, the sliding induced \PS~tends to suppress charge transfer between monolayers, leading to a smaller bandgap. Conversely, in the HP state where \PS~promotes the charge transfer, the bandgap is larger. It is important to note that the correlation between $E_g$ and the magnitude of total polarization is exactly \textit{\textbf{opposite}} to that in homobilayers of the $p+p$ configuration.

Our DFT investigations of 1H-WSe$_2$/1H-$AX_2$ ($A =$ Mo, W; $X =$ S, Se, Te) bilayers confirm the model analysis discussed above, revealing that the HP state indeed has a larger bandgap, with the values of \DEg~range from 0.09 to 0.14 eV, as shown in Fig.~\ref{fig_hetero}d. The projected density of states (PDOS) diagrams in Fig.~\ref{fig_hetero}e and f for 1H-WSe$_2$/1H-WS$_2$ heterobilayer show a clear upward shift of the CBM of the high-work-function layer in HP states, consistent with schematics in Fig.~\ref{fig_hetero}a and b.


A potential issue for a hetrobilayer is the lattice mismatch-induce formation of moir\'e patterns, which could complicate the control over sliding ferroelectricity. We suggest that doping presents a viable solution to mitigate the mismatch and enhance control. For instance, the lattice mismatch between 1H-WTe$_2$ (with an in-plane lattice constant $a_{\rm IP}^{\rm WTe_2} = $ 3.52 \AA) and 1H-WSe$_2$ ($a_{\rm IP}^{\rm WSe_2} = $ 3.30 \AA) can be significantly reduced by doping 1H-WSe$_2$ with Te atoms ($a_{\rm IP}^{\rm WSeTe} = $ 3.40 \AA). This doping approach is feasible, as our design principle does not require an ordered arrangement of Te dopants within the lattice. We perform a set of model calculations for 1H-$AX_2$/1H-$AX{\rm Te}$ ($A =$ Mo, W; $X =$ S, Se), where 1H-$AX{\rm Te}$ approximates a (homogeneously) Te-doped nonpolar monolayer (Fig.~\ref{fig_hetero}g, not a polar Janus monolayer). Similar to the undoped heterobilayer, the HP state consistently exhibits a larger bandgap than the LP state (Fig.~\ref{fig_hetero}h), further confirming the robustness of the design principle. 

\begin{figure}[t]
\centering
\includegraphics[width=13cm]{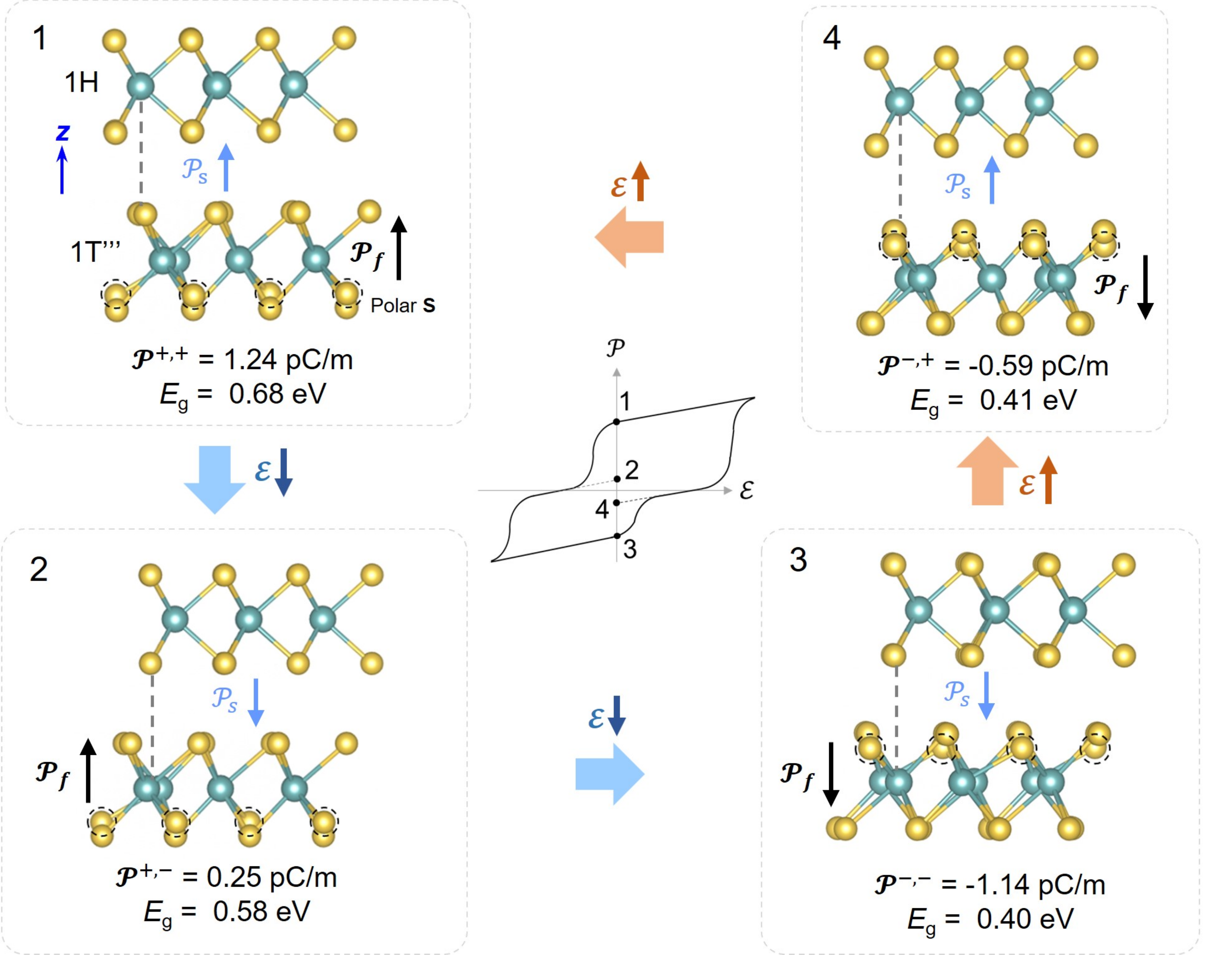}
\caption{On-demand voltage control of multiple bandgap values in a 1H-MoS$_2$/1T$'''$-MoS$_2$ heterojunction, supporting both monolayer ferroelectricity (\PF) and sliding ferroelectricty (\PS). Monolayer ferroelectricity arises from the intralayer charge transfer involving polar sulfur (S) atoms, which are highlighted within dotted circles.}
\label{fig_multi}
\end{figure}

Reversible and precisely tunable multistate switching is essential for enabling high-density information storage and neuromorphic computing applications~\cite{Li24p2653,Prosandeev22p116201}.Finally, we demonstrate that combining a 2D ferroelectric with a nonpolar monolayer (denoted as $f$+$np$) presents a compelling route to achieve fine-tuned and non-volatile control over multiple bandgaps using voltage. 
In this bilayer, there are three types of polarization: $\mathcal{P}_{f}$ originating from monolayer's ferroelectricity,  $\mathcal{P}_{\rm bg}$ due to the work function difference, and \PS~arising from sliding ferroelectricity. 
This leads to four polar states in this bilayer system: $\mathcal{P}_{\rm bg}-\mathcal{P}_{ f}-\mathcal{P}_{\rm s}$, $\mathcal{P}_{\rm bg}-\mathcal{P}_{f}+\mathcal{P}_{\rm s}$, $\mathcal{P}_{\rm bg}+\mathcal{P}_{f}-\mathcal{P}_{\rm s}$, and $\mathcal{P}_{\rm bg}+\mathcal{P}_{f}+\mathcal{P}_{\rm s}$, labeled as $\mathcal{P}^{-,-}$, $\mathcal{P}^{-,+}$, $\mathcal{P}^{+,-}$, and $\mathcal{P}^{+,+}$, respectively. We focus on a possible experimental realization of such a bilayer system, specifically a 1H-MoS$_2$/1T$'''$-MoS$_2$ heterojunction. It is noted that 1T$'''$-MoS$_2$ featuring a switchable out-of-plane polarization has been experimentally synthesized recently~\cite{Huangfu24p14708}. The minimum energy pathways for the reversal processes of \PF~and \PS~are identified using the DFT-based NEB technique, respectively. The barrier for switching \PF~is 0.043 eV/atom, higher than the 0.012 eV/atom required for switching \PS. This difference likely translates to multi-step polarization switching under an applied electric field ($\mathcal{E}$). Starting at the $\mathcal{P}^{+,+}$ state, the bandgap evolves from 0.68~eV to 0.58~eV at the partially switched $\mathcal{P}^{+,-}$ state, and then becomes 0.40~eV at the fully reversed $\mathcal{P}^{-,-}$ state (Fig.~\ref{fig_multi}). Applying an opposite electric field can recover the $\mathcal{P}^{+,+}$ state eventually. 
The multi-step polarization switching under an applied electric field, coupled to the corresponding changes in multiple bandgap values, highlights the potential for developing multistate memory devices. 

In summary, we have demonstrated a general design principle to realize reversible and non-volatile bandgap modulation by taking advantage of both sliding ferroelectricity and structural asymmetry in 2D multilayers. The irreversible background polarization arising from the structural asymmetry is modulated by the reversible polarization due to sliding, leading to polar states with different polarization magnitudes and thereby varying built-in depolarization fields that strongly affect the electronic structures. A notable advantage of this design principle is that it is applicable to both homobilayers of polar monolayers as well as heterobilayers of nonpolar monolayers. Based on experimentally synthesized 2D Janus materials, we demonstrate substantial bandgap modulations up to 0.3 eV. The introduction of ferroelectric monolayer such as 1T$'''$-MoS$_2$ into a bilayer to enable switchable background polarization allow for 
on-demand control of multiple bandgap values via voltage, offering exciting opportunities for adaptive electronics and multistate memory. \\


This work is supported by National Natural Science Foundation of China (12304128,12074319) and Westlake Education Foundation. The computational resource is provided by Westlake HPC Center. 

\bibliography{SL}
\clearpage
\newpage

\clearpage
\newpage

\clearpage
\newpage

\clearpage
\newpage

\clearpage
\newpage

\end{document}